\def\pmb#1{\setbox0=\hbox{#1}%
   \kern-.025em\copy0\kern-\wd0
   \kern.05em\copy0\kern-\wd0
   \kern-0.025em\raise.0433em\box0}
\def\gta{\mathrel{{\lower 3pt\hbox{$\mathchar"218$}}\hskip-8pt
   \raise 2pt\hbox{$\mathchar"13E$}}}
\def\lta{\mathrel{{\lower 3pt\hbox{$\mathchar"218$}}\hskip-8pt
   \raise 2pt\hbox{$\mathchar"13C$}}}
\def\half{{\scriptstyle{1\over2}}}
\def\dagg{\phantom{\dagger}}            

\def\today{\number\day\space\ifcase\month\or
  January\or February\or March\or April\or May\or June\or
  July\or August\or September\or October\or November\or December\fi
 \space\number\year}
\documentstyle[editedvolume,NATO,numreferences]{crckapb}

\font\tenrm=cmr8

\begin{opening}
\title{STRIPE-LIKE INHOMOGENEITIES, CARRIERS, AND\protect\\ 
BCS--BEC CROSSOVER IN THE HIGH-\pmb{$T_c$}\/ CUPRATES} 

\author{J. ASHKENAZI}
\institute{Physics Department, University of Miami\\
         P.O. Box 248046, Coral Gables, FL 33124, U.S.A.}

\end{opening}

\runningtitle{STRIPE-LIKE INHOMOGENEITIES\dots}

\begin{document}


\begin{abstract}
Considering both ``large-$U$'' and ``small-$U$'' orbitals, it is found
that the carriers of the high-$T_c$ cuprates are polaron-like
``stripons'' carrying charge and located on stripe-like inhomogeneities,
``quasi-electrons'' carrying charge and spin, and ``svivons'' carrying
spin and lattice distortion. This is shown to result in the observed
anomalous spectroscopic and transport properties of the cuprates.
Pairing results from transitions between pair states of stripons and
quasi-electrons through the exchange of svivons, and a crossover occurs
between BCS and Bose-Einstein condensation behaviors. 
\end{abstract}

\section{Introduction}

Theoretical calculations \cite{Zaanen,Emery1}, and a variety of
experimental data \cite{Stripes} support the assumption
that the high-$T_c$ cuprates are characterized by dynamical stripe-like
inhomogeneities, where narrow charged stripes form antiphase domain
walls separating wider antiferromagnetic (AF) stripes. Experimental
observations have been pointing to the presence of both itinerant and
almost localized (or polaron-like) carriers in the cuprates.
First-principles calculations \cite{Andersen} support an approach based
on the existence of both large-$U$ and small-$U$ orbitals in the
vicinity of the Fermi level ($E_{_{\rm F}}$). 

The fermion creation operator of a small-$U$ electron in band $\nu$,
spin $\sigma$ (which can be assigned a number $\pm 1$), and wave vector
${\bf k}$ is denoted here by $c_{\nu\sigma}^{\dagger}({\bf k})$. The
creation operators of the large-$U$ electrons in the CuO$_2$ planes are
expressed using the ``slave-fermion'' method \cite{Barnes}. Such an
electron in site $i$ and spin $\sigma$ is created by
$d_{i\sigma}^{\dagger} = e_i^{\dagger} s_{i,-\sigma}^{\dagg}$, if it is
in the ``upper-Hubbard-band'', and by $d_{i\sigma}^{\prime\dagger} =
\sigma s_{i\sigma}^{\dagger} h_i^{\dagg}$, if it is in a Zhang-Rice-type
``lower-Hubbard-band''. Here $e_i^{\dagg}$ and $h_i^{\dagg}$ are
(``excession'' and ``holon'') fermion operators, and
$s_{i\sigma}^{\dagg}$ are (``spinon'') boson operators. These auxiliary
operators have to satisfy the constraint: $e_i^{\dagger} e_i^{\dagg} +
h_i^{\dagger} h_i^{\dagg} + \sum_{\sigma} s_{i\sigma}^{\dagger}
s_{i\sigma}^{\dagg} = 1$. 

Further details on results presented here are given elsewhere
\cite{Ashk1}. 

\section{Auxiliary Space and Fields}

An auxiliary space is introduced within which a chemical-potential-like
Lagrange multiplier imposes the constraint on the average. The
projection of observable quantities to the physical space is achieved by
expressing them as combinations of Green's functions of the auxiliary
space, whose time evolution is determined by a Hamiltonian obeying the
constraint rigorously. Thus it is expected to be maintained as long as
justifiable approximations are used. 

In the zeroth order, the spinon field is diagonalized by applying the
Bogoliubov transformation \cite{Ashk2}: $s_{\sigma}^{\dagg}({\bf k}) =
\cosh{(\xi_{{\bf k}})} \zeta_{\sigma}^{\dagg}({\bf k}) +
\sinh{(\xi_{{\bf k}})} \zeta_{-\sigma}^{\dagger}(-{\bf k})$. The
``bare'' spinons, created by $\zeta_{\sigma}^{\dagger}({\bf k})$, have
energies $\epsilon^{\zeta} ({\bf k})$ with a V-shape zero minimum at
${\bf k}={\bf k}_0$. Bose condensation results in AF order at wave
vector ${\bf Q}=2{\bf k}_0$. 

Along the one dimensional charged stripes the spin-charge separation
approximation is expected to be valid. Thus it is justified there to
decouple two-particle spinon-holon (spinon-excession) Green's functions
into single-auxiliary-particle Green's functions, and to interpret the
auxiliary particles as physical quasiparticles. We refer to holons
(excessions) along the charged stripes as ``stripons'', created by
$p^{\dagger}_{\mu}({\bf k})$, having bare energies by
$\epsilon^p_{\mu}({\bf k})$, and carrying charge $-{\rm e}$. 

Since the dynamical stripe-like inhomogeneities are highly disordered,
we find it appropriate to assume zeroth order localized stripon states.
Their ${\bf k}$ wavenumbers present ${\bf k}$-symmetrized combinations
of degenerate localized states to be treated in a perturbation
expansion. Their values are within a Brillouin zone (BZ) based on
periodic supercells which are large enough to approximately contain
(each) the entire spectrum $\epsilon^p_{\mu}$ of bare stripon energies.
These supercells introduce an approximate long-range order in spite of
the local inhomogeneities. 

We do not assume spin-charge separation away from the charged stripes,
and construct approximate fermion creation operators of coupled
holon-spinon  and excession-spinon basis states as follows: 
\begin{eqnarray}
f_{\lambda\sigma}^{\dagger}({\bf k}^{\prime}, {\bf k}) &=&
e_{\lambda}^{\dagger}({\bf k}^{\prime})
s_{\lambda,-\sigma}^{\dagg}({\bf k}^{\prime} - {\bf k}) \big/
\sqrt{n^e_{\lambda}({\bf k}^{\prime}) + n^s_{\lambda,-\sigma}({\bf
k}^{\prime} - {\bf k})}, \\ 
g_{\lambda\sigma}^{\dagger}({\bf k}^{\prime}, {\bf k}) &=& \sigma
h_{\lambda}^{\dagg}({\bf k}^{\prime}) s_{\lambda\sigma}^{\dagger}({\bf
k} - {\bf k}^{\prime}) \big/ \sqrt{n^h_{\lambda}({\bf k}^{\prime}) +
n^s_{\lambda\sigma}({\bf k} - {\bf k}^{\prime})}, 
\end{eqnarray} 
where the index $\lambda$ accounts for the structure of the unit cell,
and: $n^e_{\lambda}({\bf k}) \equiv \langle e_{\lambda}^{\dagger}({\bf
k}) e_{\lambda}^{\dagg}({\bf k}) \rangle$, $n^h_{\lambda}({\bf k})
\equiv \langle h_{\lambda}^{\dagger}({\bf k}) h_{\lambda}^{\dagg}({\bf
k}) \rangle$, $n^s_{\lambda\sigma}({\bf k}) \equiv \langle
s_{\lambda\sigma}^{\dagger}({\bf k}) s_{\lambda\sigma}^{\dagg}({\bf k})
\rangle$. 

The states created by $f_{\lambda\sigma}^{\dagger}({\bf k}^{\prime},
{\bf k})$ and $g_{\lambda\sigma}^{\dagger}({\bf k}^{\prime}, {\bf k})$
are orthogonalized to the stripon states, and depleted to avoid
over-completeness. Together with the small-$U$ states [created by
$c_{\nu\sigma}^{\dagger}({\bf k})$] they form, within the auxiliary
space, a basis to ``quasi-electron'' (QE) states whose creation
operators are denoted by $q_{\iota\sigma}^{\dagger}({\bf k})$, and
mean-field bare energies by $\epsilon^q_{\iota} ({\bf k})$. These
energies form quasi-continuous ranges of bands within the BZ around
$E_{_{\rm F}}$. 

The hopping and hybridization terms of the Hamiltonian introduce strong
coupling between the QE, stripon and spinon fields. This can be
expressed through a coupling Hamiltonian term whose parameters can be in
principle derived self-consistently from the original Hamiltonian. For
the case of p-type cuprates this coupling Hamiltonian can be expressed
as:
\begin{eqnarray}
{\cal H}^{\prime} &=& {1 \over \sqrt{N}} \sum_{\iota\lambda\mu\sigma}
\sum_{{\bf k}, {\bf k}^{\prime}} \big\{\sigma
\epsilon^{qp}_{\iota\lambda\mu}({\bf k}, {\bf k}^{\prime})
q_{\iota\sigma}^{\dagger}({\bf k}) p_{\mu}^{\dagg}({\bf k}^{\prime})
[\cosh{(\xi_{\lambda,{\bf k} - {\bf k}^{\prime}})}
\zeta_{\lambda\sigma}^{\dagg}({\bf k} - {\bf k}^{\prime}) \nonumber \\
&\ &+ \sinh{(\xi_{\lambda,{\bf k} - {\bf k}^{\prime}})}
\zeta_{\lambda,-\sigma}^{\dagger}({\bf k}^{\prime} - {\bf k})] + h.c.
\big\}, 
\end{eqnarray} 
where the ${\bf k}$ values correspond to the stripons BZ, within which
the other fields have been embedded. ${\cal H}^{\prime}$ introduces a
vertex between the QE, stripon and spinon propagators \cite{Ashk3}.

A localized stripon modifies the lattice around it \cite{Bian}.
Consequently, physical process induced by ${\cal H}^{\prime}$, in which
a stripon is transformed into a QE, or vice versa, and a spinon is
emitted and/or absorbed, involves also the emission and/or absorption of
phonons. This can be described by multiplying a spinon propagator,
linked to the ${\cal H}^{\prime}$ vertex with a power series of phonon
propagators \cite{Ashk3}. Such a phonon-``dressed" spinon is referred to
as a ``svivon", carrying spin and lattice distortion, and the ${\cal
H}^{\prime}$ vertex is re-interpreted as coupling between a QE, stripon,
and svivon propagators. 

\section{Auxiliary Spectral Functions}

The electrons spectral function are expressed in terms of auxiliary
spectral functions $A^q_{\iota}({\bf k}, \omega)$, $A^p_{\mu}({\bf k},
\omega)$, and $A^{\zeta}_{\lambda}({\bf k}, \omega)$ of the QE's,
stripons, and svivons, respectively [$A(\omega) \equiv \Im {\cal
G}(\omega-i0^+) / \pi$, where ${\cal G}$ are the Green's functions].
Since the resulting stripon bandwidth is much smaller than the QE and
svivon bandwidths, a phase-space argument as in the Migdal theorem
results in negligible vertex corrections to the ${\cal H}^{\prime}$
vertex. Consequently \cite{Ashk1,Ashk3} the following expressions are
obtained for the QE, stripon, and svivon scattering rates $\Gamma^q({\bf
k}, \omega)$, $\Gamma^p({\bf k}, \omega)$, and $\Gamma^{\zeta}({\bf k},
\omega)$ [$\Gamma({\bf k}, \omega) \equiv 2\Im \Sigma({\bf k},
\omega-i0^+)$]: 
\begin{eqnarray}
\Gamma^q_{\iota\iota^{\prime}}({\bf k}, \omega) &\cong& {2\pi \over N}
\sum_{\lambda\mu{\bf k}^{\prime}} \int d\omega^{\prime}
\epsilon^{qp}_{\iota\lambda\mu}({\bf k}^{\prime}, {\bf
k})\epsilon^{qp}_{\iota^{\prime}\lambda\mu}({\bf k}^{\prime}, {\bf k})^*
A^p_{\mu}({\bf k}^{\prime}, \omega^{\prime}) \nonumber \\ &\times&
[-\cosh{^2(\xi_{\lambda,{\bf k} - {\bf k}^{\prime}})}
A^{\zeta}_{\lambda}({\bf k} - {\bf k}^{\prime}, \omega -
\omega^{\prime}) \nonumber \\ &+& \sinh{^2(\xi_{\lambda,{\bf k} - {\bf
k}^{\prime}})} A^{\zeta}_{\lambda}({\bf k} - {\bf k}^{\prime},
\omega^{\prime} - \omega)] [f_{_T}(\omega^{\prime}) +
b_{_T}(\omega^{\prime} - \omega)], \ \\ 
\Gamma^p_{\mu\mu^{\prime}}({\bf k}, \omega) &\cong& {2\pi \over N}
\sum_{\iota{\bf k}^{\prime}\sigma} \int d\omega^{\prime}
\epsilon^{qp}_{\iota\lambda\mu}({\bf k}^{\prime}, {\bf k})^*
\epsilon^{qp}_{\iota\lambda\mu^{\prime}}({\bf k}^{\prime}, {\bf k})
A^q_{\iota}({\bf k}^{\prime}, \omega^{\prime}) \nonumber \\ &\times&
[\cosh{^2(\xi_{\lambda,{\bf k}^{\prime} - {\bf k}})}
A^{\zeta}_{\lambda}({\bf k}^{\prime} - {\bf k}, \omega^{\prime} -
\omega) \nonumber \\ &-& \sinh{^2(\xi_{\lambda,{\bf k}^{\prime} - {\bf
k}})} A^{\zeta}_{\lambda}({\bf k}^{\prime} - {\bf k}, \omega -
\omega^{\prime})] [f_{_T}(\omega^{\prime}) + b_{_T}(\omega^{\prime} -
\omega)], \ \\ 
\Gamma^{\zeta}_{\lambda\lambda^{\prime}}({\bf k}, \omega) &\cong& {2\pi
\over N} \sum_{\iota{\bf k}^{\prime}\mu} \int d\omega^{\prime}
\epsilon^{qp}_{\iota\lambda\mu}({\bf k}^{\prime}, {\bf k}^{\prime} -
{\bf k})^* \epsilon^{qp}_{\iota\lambda^{\prime}\mu}({\bf k}^{\prime},
{\bf k}^{\prime} - {\bf k}) \nonumber \\ &\times&
[\cosh{(\xi_{\lambda{\bf k}})} \cosh{(\xi_{\lambda^{\prime}{\bf k}})}
A^q_{\iota}({\bf k}^{\prime}, \omega^{\prime}) A^p_{\mu}({\bf
k}^{\prime} - {\bf k}, \omega^{\prime} - \omega) \nonumber \\ &+&
\sinh{(\xi_{\lambda{\bf k}})} \sinh{(\xi_{\lambda^{\prime}{\bf k}})}
A^q_{\iota}({\bf k}^{\prime}, -\omega^{\prime}) A^p_{\mu}({\bf
k}^{\prime} - {\bf k}, \omega - \omega^{\prime})] \nonumber \\ &\times&
[f_{_T}(\omega^{\prime} - \omega)- f_{_T}(\omega^{\prime})], 
\end{eqnarray} 
where $f_{_T}(\omega)$ and $b_{_T}(\omega)$ are the Fermi and Bose
distribution functions at temperature $T$. 

A solution with low-energy singularities is obtained \cite{Ashk1} for
the intermediary energy range ($\gta 0.02\;$eV), where energies above
few tenths of an eV are cut off, introducing spurious logarithmic
divergences at $\pm\omega_c$. Expressions are derived, where the
dependencies on ${\bf k}$ and band indices are omitted for simplicity
(all the coefficients are positive, and analyticity is restored within
the low-energy range $\lta 0.02\;$eV): 
\begin{eqnarray}
A^q(\omega) &\cong& \cases{a^q_+\omega + b^q_+ \;,& for $\omega>0$,\cr
-a^q_-\omega + b^q_- \;,& for $\omega<0$,\cr} \\ 
A^p(\omega) &\cong& \delta(\omega), \\ 
A^{\zeta}(\omega) &\cong& \cases{a^{\zeta}_+\omega + b^{\zeta}_+ \;,&
for $\omega>0$,\cr a^{\zeta}_-\omega - b^{\zeta}_- \;,& for
$\omega<0$,\cr} 
\end{eqnarray} 
\begin{eqnarray}
{\Gamma^q(\omega) \over 2\pi} &\cong& \cases{c^q_+\omega + d^q_+ \;,&
for $\omega>0$,\cr -c^q_-\omega + d^q_- \;,& for $\omega<0$,\cr} \\ 
{\Gamma^p(\omega) \over 2\pi} &\cong& \cases{c^p_+\omega^3 +
d^p_+\omega^2 + e^p_+\omega \;,& for $\omega>0$,\cr -c^p_-\omega^3 +
d^p_-\omega^2 - e^p_-\omega \;,& for $\omega<0$,\cr} \\ 
{\Gamma^{\zeta}(\omega) \over 2\pi} &\cong& \cases{c^{\zeta}_+\omega +
d^{\zeta}_+ \;,& for $\omega>0$,\cr c^{\zeta}_-\omega - d^{\zeta}_- \;,&
for $\omega<0$,\cr} 
\end{eqnarray} 
\begin{eqnarray} 
-\Re\Sigma^q(\omega) &\cong& \omega_c(c^q_+ - c^q_-) + \big(d^q_+
\ln{\big|{\omega - \omega_c \over \omega}\big|} - d^q_- \ln{\big|{\omega
+ \omega_c \over \omega}\big|}\big) \nonumber \\ &+& \omega\big(c^q_+
\ln{\big|{\omega - \omega_c \over \omega}\big|} + c^q_- \ln{\big|{\omega
+ \omega_c \over \omega}\big|}\big), \\ 
-\Re\Sigma^p(\omega) &\cong& \big[{\omega_c^3 \over 3}(c^p_+ - c^p_-) +
{\omega_c^2 \over 2}(d^p_+ - d^p_-) + \omega_c(e^p_+ - e^p_-)\big]
\nonumber \\ &+& \omega \big[{\omega_c^2 \over 2}(c^p_+ + c^p_-) +
\omega_c(d^p_+ + d^p_-) + e^p_+ \ln{\big|{\omega - \omega_c \over
\omega}\big|} \nonumber \\ &+& e^p_- \ln{\big|{\omega + \omega_c \over
\omega}\big|}\big] + \omega^2 \big[\omega_c(c^p_+ - c^p_-) + d^p_+
\ln{\big|{\omega - \omega_c \over \omega}\big|} \nonumber \\ &-& d^p_-
\ln{\big|{\omega + \omega_c \over \omega}\big|}\big] + \omega^3
\big[c^p_+ \ln{\big|{\omega - \omega_c \over \omega}\big|} + c^p_-
\ln{\big|{\omega + \omega_c \over \omega}\big|}\big], \\ 
-\Re\Sigma^{\zeta}(\omega) &\cong& \omega_c(c^{\zeta}_+ + c^{\zeta}_-) +
\big(d^{\zeta}_+ \ln{\big|{\omega - \omega_c \over \omega}\big|} +
d^{\zeta}_- \ln{\big|{\omega + \omega_c \over \omega}\big|}\big)
\nonumber \\ &+& \omega\big(c^{\zeta}_+ \ln{\big|{\omega - \omega_c
\over \omega}\big|} - c^{\zeta}_- \ln{\big|{\omega + \omega_c \over
\omega}\big|}\big). 
\end{eqnarray} 
The auxiliary-particle energies are renormalized: $\bar\epsilon =
\epsilon + \Re\Sigma(\bar\epsilon)$. This renormalization is
particularly strong on the stripon energies where the bandwidth drops
down to the low energy range. 

For the discussed case of p-type cuprates the following inequalities
between the coefficients in Eqs.~(7--12) are generally expected
\cite{Ashk1}: 
\begin{eqnarray}
&\ &a^q_+>a^q_-,\ \ \ \  b^q_+>b^q_-,\ \ \ \  c^q_+>c^q_-,\ \ \ \ 
d^q_+>d^q_-, \  \\ &\ &a^{\zeta}_+>a^{\zeta}_-,\ \ \ \ 
b^{\zeta}_+>b^{\zeta}_-,\ \ \ \ c^{\zeta}_+>c^{\zeta}_-,\ \ \ \ 
d^{\zeta}_+>d^{\zeta}_-. \  
\end{eqnarray} 
For ``real'' n-type cuprates (namely, ones where the stripons are based
on excession and not holon states) the direction of the inequalities is
reversed for the QE coefficients (16). 

\section{Consequences for Electron Spectroscopies}

In order to apply the auxiliary spectral functions in Eqs.~(7--9) for
physical properties, a projection into the physical space is required.
The low-energy signature of the stripon spectral functions is not
detected by spectroscopies, like ARPES, which measure the effect of
transfer of electrons into, or out of, the crystal. It is smeared over
few tenths of an eV through convolution with svivon spectral functions,
and contributes an ``incoherent'' background. On the other hand,
transport properties measure the electrons {\it within} the crystal, and
can detect the stripon energy scale. The physical contribution of the
quasi-continuum of QE bands to the electron spectral functions is strong
only for few QE bands for which the expansion coefficients in terms of
their basis states [created by (1,2) $f_{\lambda\sigma}^{\dagger}({\bf
k}^{\prime}, {\bf k})$, $g_{\lambda\sigma}^{\dagger}({\bf k}^{\prime},
{\bf k})$, and $c_{\nu\sigma}^{\dagger}({\bf k})$] closely correspond to
those of real electron states, and they contribute ``coherent''
$\bar\epsilon^q ({\bf k})$ bands. The other QE states also contribute to
the incoherent background. The dependence of the QE basis states (1,2)
on $n^e_{\lambda}$, $n^h_{\lambda}$, and $n^s_{\lambda\sigma}$ is
reflected in the observed dependence \cite{Eskes} of strongly correlated
electrons spectrum on these occupation factors. 

The characteristics of the auxiliary spectral functions are maintained in
their projection to the physical space (detected, {\it e.g.}, by ARPES).
The fact that the QE field is coupled to the stripon and svivon fields
is reflected, {\it e.g.}, in ``shadow bands''. The results for
$\Gamma^q$ (10) are reflected in the observed non-Fermi-liquid
bandwidths, having a $\propto\omega$ and a constant term. The results
for $\Re\Sigma^q$ (13) are reflected in the renormalization of band
slopes, compared to LDA predictions. The consequence of the logarithmic
singularity in $\Re\Sigma^q$ at $\omega=0$ [due to the $(d^q_+ -
d^q_-)\ln{|\omega|}$ term in Eq.~(13)], for p-type cuprates (16), is
that as an $\bar\epsilon^q({\bf k})$ band approaches $E_{_{\rm F}}$ from
below it is becoming flatter (though it does not narrow accordingly)
while from above it becomes steeper, and may pass through an infinite
slope resulting in an S-shape triply valued band (causing smearing of
the spectral weight, and blurring the band). The effect of the
singularity is truncated and smoothened on an energy scale $\lta
0.02\;$eV. For ``real'' n-type cuprates these behaviors below and above
$E_{_{\rm F}}$ are expected to switch. ARPES data reflects the behavior
below $E_{_{\rm F}}$, and results of such band-flattening have been
reported in p-type cuprates (and attributed to electron coupling with
phonons \cite{Shen} or with the neutron scattering resonance mode
\cite{Johnson}). The behavior above $E_{_{\rm F}}$ would be detected in
ARIPES measurements. Further measurements of the bands of n-type
cuprates very close to $E_{_{\rm F}}$ are necessary to determine their
behavior there. 

The renormalization of svivon energies (from $\epsilon^{\zeta}$ to
$\bar\epsilon^{\zeta}$) due to $\Re\Sigma^{\zeta}(\omega)$ (15), and
specifically its logarithmic singularity at $\omega=0$ due to the
$(d^{\zeta}_+ + d^{\zeta}_-)\ln{|\omega|}$ term (which is smoothened at
low energies) results in a negative shift close to the minimum at ${\bf
k}_0$; and as ${\bf k}_0$ is approached, the slope of
$\bar\epsilon^{\zeta}({\bf k})$ increases first [compared to the almost
constant slope of $\epsilon^{\zeta}({\bf k)}$] and it may become
infinite, resulting in a range of an S-shape triply valued and blurred
band; but as ${\bf k}_0$ is further approached, the slope of
$\bar\epsilon^{\zeta}({\bf k})$ decreases and becomes smaller than the
slope of $\epsilon^{\zeta}({\bf k})$ at its minimum at ${\bf k}_0$. From
analyticity arguments (due to the absence of long-range AF order) it is
expected that $\bar\epsilon^{\zeta}({\bf k})$ has a smooth minimum,
rather than the V-shape minimum of $\epsilon^{\zeta}({\bf k})$. Thus
$-\bar\epsilon^{\zeta}({\bf k}_0)$ presents the crossover energy from
the intermediary to the low energy range. Since spin-flip excitations
are, largely, double-svivon absorption or emission processes,
$-2\bar\epsilon^{\zeta}({\bf k}_0)$ is expected to be the energy of such
an excitation at ${\bf k} = {\bf Q}$. This result is consistent with the
neutron-scattering resonance found at this wave number \cite{Fong}. It
is observed that this resonance is at a local energy maximum, for ${\bf
k}$ around ${\bf Q}$, consistently with our result that
$\bar\epsilon^{\zeta}({\bf k}_0)$ is at a minimum below zero. 

An anomalous optical conductivity has been observed for the high-$T_c$
cuprates \cite{Tanner}, consisting of Drude and mid-IR terms. This is
consistent with our results, where the Drude term is due to transitions
between QE states, and between stripon states, while the mid-IR term
largely results from transitions between stripon states and QE+svivon
states. The mid-IR term becomes negligibly small for $\omega \to 0$,
where the conductivity could be decoupled into QE and stripon terms. 

\section{Transport Properties}

Normal-state transport expressions [{\it not} including the effect of
the pseudogap (PG)] are obtained \cite{Ashk1} using linear response
theory, where the zero-energy singularities in the auxiliary spectral
functions and (7--9) are smoothened in the low-energy scale [and
$A^{\zeta}(\omega=0)=0$]. Also it is taken into account that the
electric current can be expressed as a sum ${\bf j} = {\bf j}^q_0 + {\bf
j}^p_0$ of contributions of ``bare'' QE and and stripon states, where
${\bf j}^p_0 \cong 0$ since the bare stripon states are localized. 

Results for the electrical resistivity $\rho$, the Hall constant
$R_{_{\rm H}}$, the Hall number $n_{_{\rm H}} = 1/{\rm e} R_{_{\rm H}}$,
the Hall angle $\theta_{_{\rm H}}$ (through $\cot{\theta_{_{\rm H}}} =
\rho / R_{_{\rm H}}$), and the thermoelectric power (TEP) $S$, are
presented in Fig.~1. Their anomalous temperature dependencies result
both from the low energy scale stripon band (8), modeled here by a
``rectangular" shape $A^p$ of width $\omega^p$ and fractional occupation
$n^p$, and from those of the scattering rates $\Gamma^q(T,\omega=0)$
and $\Gamma^p(T,\omega=0)$, derived using Eqs.~(4,5) to which
temperature independent terms are added to account for impurity
scattering. 

\begin{figure}[t]
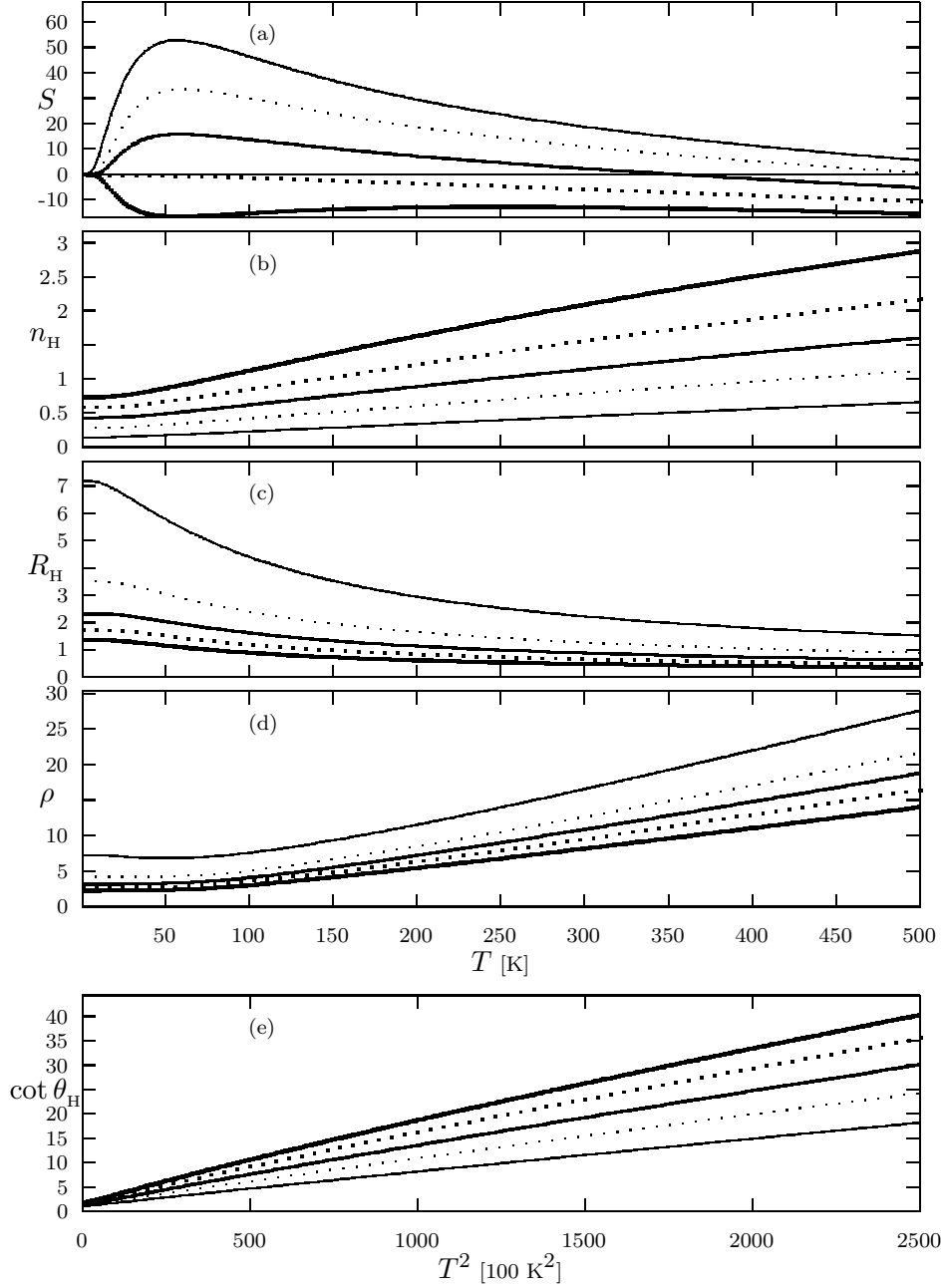


\setlength{\unitlength}{0.240900pt}
\ifx\plotpoint\undefined\newsavebox{\plotpoint}\fi
\sbox{\plotpoint}{\rule[-0.200pt]{0.400pt}{0.400pt}}%


\caption{The transport coefficients, in arbitrary units [and $\mu$V/K
units for $S$ (a)], for: $n^p$=0.8,0.7,0.6,0.5,0.4;
$10000N^q_e$=20,23,26,29,32; $\omega^p$[K]=200,190,180,170,160;
$n^p_{_{\rm H}}$=0.1,0.2,0.3,0.4,0.5; $n^q_{_{\rm H}}$=6,7,8,9,10;
$S^q_1$=$-$0.025; $\gamma^p_0$=500; $\gamma^p_2$=0.03; $\gamma^q_0$=5;
$\gamma^q_1$=0.2. The last values correspond to the thickest lines.} 
\label{F1}
\end{figure}

The transport results in Fig.~1 correspond to five stoichiometries,
ranging from underdoped ($n^p=0.8$) to overdoped ($n^p=0.4$) p-type
cuprates. The parameter $N^q_e$, corresponding to the QE contribution to
the electrons density of states at $E_{_{\rm F}}$, is assumed to
increase with doping, reflecting transfer of QE spectral weight towards
$E_{_{\rm F}}$. Consequently (5) $\omega^p$ is assumed to decrease with
doping. The parameters $n^q_{_{\rm H}}$ and $n^p_{_{\rm H}}$ represent
effective QE and stripon contributions to the density of charge carriers
(reflected in the Hall number). Since they both contribute through the
current ${\bf j}^q_0$ of the bare QE states, they are expected to have
same sign (corresponding to these states). The values of these
parameters are assumed to increase with doping, where $n^p_{_{\rm H}}$
increases considerably faster than $n^q_{_{\rm H}}$, and a little faster
than $1-n^p$ (since $n^p_{_{\rm H}}$ reflects an overall stripon-related
carriers density, while $n^p$ is the fractional occupation of stripon
states within the charged stripes whose number increases with doping).
Doping-independent values are taken for the QE TEP parameter $S^q_1$
[which is negative for p-type cuprates by the inequality (16)], and for
the stripon and QE scattering rate parameters $\gamma^p_0$,
$\gamma^p_2$, $\gamma^q_0$, and $\gamma^q_1$. 

The TEP results depend strongly on $n^p$, and reproduce very well the
doping-dependent experimental behavior \cite{Fisher,Tanaka,Matsuura}.
The position of the maximum in $S$ depends on the choice of $\omega^p$,
and it may occur below or above $T_c$ (a PG may shift it to a higher
temperature than predicted here). Also the other transport coefficient
in Fig.~1 reproduce very well the experimental behavior
\cite{Kubo,Takagi}. Cases where the linear $T$ dependence of the
resistivity persists at very low temperatures are expected as a PG
effect. 

The TEP in ``real'' n-type cuprates is expected to behave similarly to
p-type cuprates, but with an opposite sign and slope. Results for NCCO
\cite{Takeda} reveal that for doping levels where it is superconducting 
(SC), the TEP slope actually corresponds to those of p-type cuprates
[Fig.~1(a)], but with an opposite effect of doping. Thus $S$ for lower
doping levels is typical of the p-type overdoped regime. This points to
the possibility that NCCO is not a ``real'' n-type cuprate, and its
stripons are also based on holon states, where the extra doped negative
charge occupies another QE band. The Hall constant in NCCO \cite{Takeda}
is also consistent with this scenario and our results. Another
possibility is that the TEP slope in NCCO is determined by electrons
which are not strongly coupled to the CuO$_2$ planes. 

\section{Mechanism --- BCS--BEC Crossover}

The electronic structure of the cuprates discussed here provides SC
pairing due to transitions between pair states of stripons and QE's
through the exchange of svivons \cite{Ashk3}. Since such transitions
enable long-range hopping of local stripon pairs, without an associated
hopping of svivons (necessary for single-stripon hopping), this type of
pairing results in a large gain in the ``kinetic'' energy of the
stripons. Thus this can effectively be expressed as a strong attractive
stripon-stripon interaction. Thus, pair-breaking is then expected to
result in both QE and stripon+svivon excitations. This is confirmed by
ARPES results in the SC state \cite{Norman}, showing a sharp peak at
$\sim 0.04\;$eV over a wide range of the BZ, and a ``hump'' starting at
$\sim 0.1\;$eV and merging with a normal-state band at higher energies,
corresponding to these excitations. The hump is consistent with the QE
pair-breaking excitation, and the peak with the stripon+svivon
pair-breaking excitation at the svivon minimum
$\bar\epsilon^{\zeta}({\bf k}_0)$. In the SC state this minimum is
within the gap, and thus is considerably narrower than in the normal
state [reflected also in the width of the neutron-scattering resonance
energy $-2\bar\epsilon^{\zeta}({\bf k}_0)$]. 

When the pairing energy is comparable to the relevant bandwidth, a
crossover is expected \cite{Randeria} to occur between BCS and
preformed-pairs Bose-Einstein condensation (BEC) behaviors. This
condition is fulfilled here due to the small stripon bandwidth, and the
underdoped cuprates are characterized by BEC behavior, where singlet
pairs are formed at $T_{\rm pair}$, and SC occurs below $T_{\rm coh} (<
T_{\rm pair})$ where phase coherence sets in. The normal-state PG is a
pair-breaking gap at $T_{\rm coh} < T < T_{\rm pair}$, having
similarities to the SC gap, and accounting for most of the pairing
energy, as observed \cite{Moca}. The coherence temperature can be
expressed \cite{Emery2} as: $T_{\rm coh} \propto n_s / m_s^*$, where
$m_s^*$ and $n_s$ are the pairs effective mass and density, in agreement
with the ``Uemura plots''. Since we expect [Fig.~1(a)] the stripon band
to be half full ($n^p=\half$) for slightly overdoped cuprates, the
``boomerang-type'' behavior \cite{Niedermayer} of these plots is
understood as a crossover between a stripon band-top $T_c = T_{\rm coh}$
behavior in the underdoped cuprates, and a stripon band-bottom $T_c =
T_{\rm pair}$ behavior in overdoped cuprates. 

\section{Conclusions}

It was shown that a treatment of the cuprates on the basis of both
large-$U$ and small-$U$ orbitals and dynamical stripe-like
inhomogeneities resolves puzzling normal-state spectroscopic and
transport properties and provides a mechanism for high-$T_c$ and the
peudogap.

\end{document}